%% file: main.tex
\documentclass[letterpaper, 10 pt, conference]{ieeeconf}

\usepackage{amsmath,amsthm,amssymb}
\usepackage{graphics}
\usepackage{hyperref}
\usepackage{etoolbox}
\usepackage{optidef}
\usepackage[utf8]{inputenc}
\usepackage{caption,subcaption}
\usepackage{pgfplots}
\usepackage{cite}
\pgfplotsset{compat=newest}
\usepgfplotslibrary{groupplots}
\usepgfplotslibrary{dateplot}
\appto\UrlBreaks{\do\-}

\IEEEoverridecommandlockouts  
 \usepackage{stackengine}
\usepackage{optidef}
\pgfplotsset{every tick label/.append style={font=\footnotesize}}
\usepackage{booktabs}
\usepackage{accents}
\usetikzlibrary{arrows,automata}
\newcommand{\ubar}[1]{\underaccent{\bar}{#1}}
\usepackage[ruled,vlined, linesnumbered]{algorithm2e}
\SetKwInput{KwInput}{Input}                
\SetKwInput{KwOutput}{Output}      
\newtheorem{definition}{Definition}
\newtheorem{lemma}{Lemma}
\newtheorem{theorem}{Theorem}

\usepackage{epsfig} 
\usepackage{times} 

\title{\LARGE \bf
Privacy-Preserving Policy Synthesis in Markov Decision Processes
}

\author{Parham Gohari, Matthew Hale and Ufuk Topcu
\thanks{This work was supported by the AFOSR Center of Excellence on Assured Autonomy in Contested Environments. MH was supported by NSF CAREER Grant 1943275.}
\thanks{P. Gohari is with the Department of Electrical and Computer Engineering,
        The University of Texas at Austin, Austin, TX. U. Topcu is with Faculty of Oden Institute for Computational Engineering, The University of Texas at Austin,
        Austin, TX.
        email: {\tt\small \{pgohari, utopcu\}@utexas.edu}. M. Hale is with Faculty of Mechanical and Aerospace Engineering at the University of Florida, Gainseville, FL. email: {\tt\small matthewhale@ufl.edu}.}%
}

\begin{document}

\maketitle
\thispagestyle{empty}
\pagestyle{empty}

\begin{abstract}

In decision-making problems, the actions of an agent may reveal sensitive information that
drives its decisions. 
For instance, a corporation's investment decisions may reveal its sensitive
knowledge about market dynamics. To prevent this type of information leakage, 
we introduce a policy synthesis algorithm that protects the privacy of the transition
probabilities in a Markov decision process.
We use differential privacy as the mathematical definition of privacy.
The algorithm first perturbs the transition probabilities using a mechanism that provides differential privacy. Then, based on the privatized transition probabilities, we synthesize a policy using dynamic programming.
Our main contribution is to bound the ``cost of privacy,'' \textit{i.e.}, the difference between
the expected total rewards with privacy and the expected total rewards without privacy. 
We also show that computing the cost of privacy has time complexity that is polynomial in the parameters of the problem. 
Moreover, we establish that the cost of privacy increases with the strength of differential privacy protections, and we quantify this increase. 
Finally, numerical experiments on two example environments validate the established relationship between the cost of privacy and the strength of data privacy protections.
\end{abstract}

\section{introduction}

In many decision-making problems, agents desire to protect sensitive information that drives their actions from eavesdroppers and adversaries, such as applications in autonomous driving or smart power grids \cite{c18, c19}.
In these applications, as well as in many other sequential decision-making problems, choosing actions can be cast as a policy-synthesis problem wherein the environment is modeled as a Markov decision process (MDP) \cite{c20, c21}.
The goal in a policy-synthesis problem is to find a reward-maximizing control policy based on the transition probabilities of the underlying MDP.
In this work, we study the problem of synthesizing a policy that protects the privacy of the transition probabilities.\par

Transition probabilities in an MDP govern the dynamics of the environment and may carry information that should be protected during policy synthesis.
For example, suppose that through market research, a corporation discovers a niche in the market and decides to invest.
Such an investment may alert competitors to the discovered niche and leads to other firms making similar decisions.
Previous works in economics have associated higher market shares with profitability \cite{c1,c2}.
Therefore, competitors' entrance to the market may be harmful to the investing corporation.
As a result, it is often crucial for a decision-maker to choose actions that do not reveal its knowledge about its environment dynamics. \par

We use differential privacy as the definition of privacy for an MDP's transition probabilities.
Differential privacy, first introduced in \cite{c3}, is a property of an algorithm and has been used in the computer science literature as a quantitative definition of privacy for databases \cite{c10, c11}.
It has also recently been used in control theory \cite{c4, c5}.
Differential privacy makes it unlikely that the output of a differentially private algorithm will reveal any useful information about the individual entries of the input dataset; however, it may still pass on information about the aggregate statistics of the input dataset that are useful in down-stream analytics.\par

The main contribution of this paper is to develop a policy-synthesis algorithm that enforces differential privacy for transition probabilities with adjustable privacy and utility.
We define the utility of a privacy-preserving policy synthesis algorithm to be the value function associated with the policy, which in an MDP is its expected total reward \cite{c9}.
Utility loss due to privacy is a common phenomenon, and we follow the convention in the differential privacy literature to analyze the utility of the privacy-preserving algorithm by comparing it to its non-private counterpart \cite{c6, c7}. \par

In order to show that the algorithm enforces differential privacy, we exploit the fact that differential privacy is immune to post-processing \cite{c6}. 
By immunity to post-processing, we mean that arbitrary functions of the output of a differentially private algorithm do not weaken its privacy guarantees. 
The algorithm first privatizes the transition probabilities via the Dirichlet mechanism \cite{c8}.
We then use dynamic programming to synthesize a policy based on the privatized transition probabilities.
Since the dynamic programming stage is an act of post-processing on the output of a differentially private mechanism, its output preserves the differential privacy provided to transition probabilities. \par

We employ the Dirichlet mechanism for privatization because it preserves the unique structure of the transition probabilities, \textit{i.e.}, vectors with non-negative components that sum to one.
Using traditional differentially private mechanisms that add infinite-support noise to transition probabilities are ill-suited to this work as they break the structure of transition probabilities.
For example, they can result in a transition probability vector with negative components.
Although normalization may seem a fitting solution in order to project the perturbed vector back onto the unit simplex, we avoid normalization because it makes it difficult to quantify utility.\par

We introduce the ``cost of privacy'' as a measure of the utility of the algorithm.
We define the cost of privacy to be the difference between the expected total rewards of the policy with privacy and that of the same policy without privacy.
Since we perturb the transition probabilities to enforce differential privacy, the output of the dynamic-programming stage is susceptible to suboptimality, which the cost of privacy quantifies. \par

We bound the cost of privacy for both finite- and infinite-horizon MDPs.
For finite-horizon MDPs, we show that we can compute the cost of privacy in polynomial time via a backward-in-time recursive algorithm.
For the case of infinite-horizon MDPs, we show that an algorithm similar to policy evaluation converges to the cost of privacy asymptotically.
We further show that the number of iterations required to approximate the cost of privacy is polynomial in problem parameters.
This work enables a decision-maker to control the level of privacy based on the utility loss that they are willing to tolerate.\par

In order to empirically validate the expressions that we introduce for the cost of privacy, we run the algorithm on two example MPDs.
The first example is a small MDP that models a corporation's investment.
The second example is an MDP with a larger state and action space, with its transition probabilities generated randomly.
We run the algorithm at a range of privacy levels and visualize our results by plotting the cost of privacy versus privacy level.
The results illustrate the trade-off that we establish between the strength of data privacy and utility.
Furthermore, we observe that the bounds we provide for the cost of privacy are meaningful in the sense that they empirically provide a close approximation to the cost of privacy. \par

\textbf{Related work. }The works in \cite{c12, c13, c14} study the problem of learning a policy in an MDP while enforcing differential privacy.
The key difference between this paper and the works above is that we protect the transition probabilities which belong to the probability simplex, whereas the other works protect the sensory data that are scalars.
We emphasize that although scalars can be readily privatized using traditional differentially private mechanisms, transition probabilities need to be treated specially to ensure that they remain non-negative and sum to one.

The problems of robust and distributionally robust MDPs are related to this paper.
Robust policy synthesis in an MDP is the problem of synthesizing a policy that mitigates uncertainties present in transition probabilities \cite{c15, c16}.
Distributionally robust MDPs assume that the planner has access to a probability measure over the uncertainty sets \cite{c17}.\par
We base the cost of privacy bounds on a concentration bound that we derive for the output of the Dirichlet mechanism.
Finding the worst-case cost of privacy coincides with lower bounding the value of a distributionally robust policy where the uncertainties in the transition probabilities adhere to the concentration bound of the Dirichlet mechanism.
Despite the fact that the solutions take similar forms, this paper studies the value loss due to privacy, whereas the problem of robust policy synthesis tries to compute a policy that mitigates the effect of uncertainties. \par

\section{preliminaries} \label{sec:prelims}

In this section we set the notation and definitions used throughout the paper.

\subsection{Notation}

We denote the set of real numbers by $\mathbb{R}$. Let $(\cdot)^T$ denote the transpose of a vector. We define the unit simplex to be $\Delta(n) := \{x\in \mathbb{R}^n \mid \textbf{1}^T x = 1, x\geq 0\}$, where $\mathbf{1}$ is the vector of all ones in $\mathbb{R}^n$ and the inequality is evaluated element wise. 
We use the notation $\Delta^\circ(n)$ to denote the interior of $\Delta(n)$.
For a finite set $\mathcal{A}$, its cardinality is denoted by $\left|\mathcal{A}\right|$. 
$\mathbf{E}[\cdot]$ and $\mathop{\text{Var}}(\cdot)$ denote the expectation and the variance of a random variable, respectively.
$\|\cdot\|_1$ and $\|\cdot\|_\infty$ denote the one and infinity norm of a vector, respectively.
For a vector $p$, we use the notation $p_i$ to denote the $i^{\text{th}}$ component of $p$.
We use the gamma function
\begin{equation*}
    \Gamma(z) := \int_0^\infty x^{z-1} \exp{(-x)}\text{d}x.
\end{equation*}

\subsection{Markov decision processes}
An MDP is a tuple $\mathcal{M}=\left(\mathcal{S}, \mathcal{A}_s, r, \mathcal{P}, T, \gamma\right)$ where $\mathcal{S}$ is the set of states, $\mathcal{A}_s$ is the set of available actions at state $s\in\mathcal{S}$, and $r:\mathcal{S}\times\mathcal{A}_s \mapsto \mathbb{R}$ is the reward function that indicates the one-step reward for taking action $a$ at state $s$.
$\mathcal{P} := \left\{P(s,a) \in \Delta(\left|\mathcal{S}\right|) \mid (s,a)\in\mathcal{S}\times\mathcal{A}_s\right\}$ is the set of transition probabilities.
Finally, $T$ is the time horizon and $\gamma$ is the discount factor.\par

We now define a policy, that is, a rule for making a sequence of decisions in an MDP. 
In particular, let $h_t := \{s_0,a_0,r_0,s_1,a_1,r_1,\dots,s_t\}$ be a history until stage $t$, and let $\mathcal{H}_t(s_t)$ denote the set of all possible histories that end in state $s_t$.
A policy $\pi: \mathcal{H}_t(s_t) \mapsto \Delta\left(|\mathcal{A}_{s_t}|\right)$ maps a history $h_t$ to a probability distribution over the set of actions, $\mathcal{A}_{s_t}$.\par

A policy $\pi$ is evaluated by its value function $V_t^\pi : \mathcal{S} \mapsto \mathbb{R}$, that is defined as
\begin{equation*}
    V^\pi_t (s) := \mathbf{E}\left[\sum\limits_{i=t}^T \gamma^{i-t} r_i \,\middle\vert\, s_t = s\right].
\end{equation*}
The expectation is taken over the stochasticity of the policy $\pi$ and transition probabilities $\mathcal{P}$.
We study the problem of privacy-preserving policy synthesis, and in a synthesis problem, the goal is to find an optimal policy, in the sense that it achieves the highest value function beginning at initial state $s_0$. \par
In this paper, we restrict our attention to Markovian policies, \textit{i.e.}, the class of policies that only depend on the most recent state of the history.
Markovian policies are shown to be optimal under some mild conditions \cite{c23}.
We use the notation $\pi_t(a \!\mid\! s)$ to show the probability of taking action $a$ at state $s$ and stage $t$.

\subsection{Differential privacy}
For an algorithm that satisfies differential privacy, it is unlikely to tell apart \textit{nearby} input datasets based on observations of the algorithm's output. 
Nearby datasets are defined formally by an adjacency relationship.
We first state the adjacency relationship used in this paper.
\begin{definition} [From \cite{c8}, Definition 1]\label{def:adjacency}
For a constant $b \in (0,1]$, two vectors $p,q \in \Delta(n)$ are said to be $b$-adjacent if there exist indices $i,j$ such that
    \begin{equation*}
        p_{-(i,j)} = q_{-(i,j)}\ \text{and} \ ||p-q||_1 \leq b.
    \end{equation*}
\end{definition}\par
The above definition considers two vectors in the unit simplex adjacent if they only differ in two indices, $i,j$, by no more than $b$ in their 1-norm.
Note that the usual adjacency relationship in the differential privacy literature considers two input datasets adjacent if they only differ in one entry \cite{c6}; however, it is not possible for the elements of the unit simplex to differ in only one entry because their components must sum to one.
\begin{definition} [Probabilistic differential privacy \cite{c8}]
Fix a probability space $(\Omega, \mathcal{F}, \mathbb{P})$ and $b\in (0,1]$. A mechanism $\mathcal{M}: \Delta(n) \times \Omega \mapsto \Delta(n)$ is said to be probabilistically $(\epsilon, \delta)$-differentially private if, for all $p \in \Delta(n)$, we can partition the output space $\Delta(n)$ into two disjoint sets, $\Omega_1, \Omega_2$, such that $\mathbb{P}[\mathcal{M}(p) \in \Omega_2] \leq \delta$, and for all $q\in\Delta(n)$ $b$-adjacent to $p$, we have that 
\begin{equation*}
    \log\left(\frac{\mathbb{P}[\mathcal{M}(p) =x]}{\mathbb{P}[\mathcal{M}(q) = x]}\right) \leq \epsilon, \forall x\in\Omega_1.
\end{equation*}
\end{definition}
Probabilistic $(\epsilon, \delta)$-differential privacy is known to imply ordinary $(\epsilon,\delta)$-differential privacy \cite{c22}.

\subsection{The Dirichlet mechanism}
A Dirichlet mechanism with parameter $k > 0$ takes as input a vector $p\in\Delta_n^\circ$ and outputs $x \in \Delta(n)$ according to a Dirichlet probability distribution. 
Fix $k$ and let $\mathcal{M}_D^{(k)}$ denote the Dirichlet mechanism.
Then
\begin{equation*}
    \mathbb{P}\left[\mathcal{M}_D^{(k)}(p)\! = \! x\right] = {\frac{1}{\text{B}(kp)}} \prod\limits_{i=1}^{n-1} x_i^{kp_i-1}\left(1-\sum\limits_{i=1}^{n-1}x_i\right)^{kp_n-1}\!,
\end{equation*}
where
\begin{equation*}
    \text{B}(kp) := \frac{\prod\limits_{i=1}^{n} \Gamma(kp_i)}{\Gamma\left(k\sum\limits_{i=1}^n p_i\right)}
\end{equation*}
is the multi-variate beta function. \par
The Dirichlet mechanism satisfies probabilistic $(\epsilon, \delta)$-differential privacy \cite{c8}, and has the following properties.
The expected value of the output is equal to the input vector, \textit{i.e.}, $\mathbf{E}\left[\mathcal{M}_D^{(k)}(p)\right] = p$.
An increase in $k$ results in weaker differential privacy protections, and in particular it increases $\epsilon$. 
However, as $k$ increases, the output becomes more concentrated around the input vector $p$.
\section{privacy-preserving synthesis algorithm}\label{sec:algorithm}
In this section, we first present the proposed privacy-preserving synthesis algorithm.
Then, we show the differential privacy of the algorithm.
\subsection{Algorithm}
\begin{algorithm}[t] 
\DontPrintSemicolon
  \KwInput{$\left(\mathcal{S}, \mathcal{A}_s, r, \mathcal{P}, T, \gamma\right)$, k}
  \KwOutput{$\bar\pi$, $\bar V^{\bar\pi}$}
  Construct the set of privatized transition probabilities $\mathcal{\bar P}:=\left\{\bar P(s,a) = \mathcal{M}_D^{(k)}\left(P(s,a)\right) \mid P(s,a) \in \mathcal{P}\right\}$.\;
  Replace $\mathcal{M}$ with its privatized version $\mathcal{\bar M} := \left(\mathcal{S}, \mathcal{A}_s, r, \mathcal{\bar P}, T, \gamma\right)$\;
  Synthesize policy $\bar \pi$ for $\mathcal{\bar M}$.\;
  Compute the value function of $\bar \pi$, $\bar V^{\bar \pi}$.\;
\caption{Privacy-preserving synthesis algorithm \label{Alg:section3partA}}
\end{algorithm}

The algorithm takes as input an MDP representation $\mathcal{M} = \left(\mathcal{S}, \mathcal{A}_s, r, \mathcal{P}, T, \gamma\right)$ and the value of $k$ that is the parameter for the Dirichlet mechanism.
It outputs a policy $\bar \pi$ and its value function $\bar V^{\bar \pi}$.
The algorithm comprises two stages.
The first stage privatizes the transition probabilities by applying the Dirichlet mechanism independently on each transition probability vector in $\mathcal{P}$.
Let $\mathcal{\bar P}:=\left\{\bar P(s,a) = \mathcal{M}_D^{(k)}\left(P(s,a)\right) \mid P(s,a) \in \mathcal{P}\right\}$ be the set of transition probabilities after privatization.
The second stage finds an optimal policy and the optimal value of the privatized MDP $\mathcal{\bar M} := \left(\mathcal{S}, \mathcal{A}_s, r, \mathcal{\bar P}, T, \gamma\right)$.
An optimal policy is one that satisfies the Bellman condition of optimality, and the optimal value is the value of such policies \cite{c23}.
In the case of a finite-horizon MDP, the second stage finds $(\bar \pi, \bar V_t)$ such that for all $t\in\{0,\dots,T-1\}$ and all $s \in \mathcal{S}$,
\begin{equation*} \label{eq:sec3subAeq1}
    \bar V_t (s) \!=\! \max_{\pi}\!\!\sum\limits_{a \in \mathcal{A}_s} \!\!\!\pi(a\!\mid\! s) \! \left(\!r(s,a) \!+\! \gamma\! \sum\limits_{s' \in \mathcal{S}} \!\Bar{P}(s,a,s')\Bar{V}_{t+1}(s') \!\!\right)\!\!,
\end{equation*}
\begin{equation*} \label{eq:sec3subAeq2}
    \bar\pi_t  \!\in\! \text{arg} \max\limits_\pi\!\!\sum\limits_{a \in \mathcal{A}_s} \!\!\!\pi(a\!\mid\! s) \! \left(\!r(s,a) \!+\! \gamma\! \sum\limits_{s' \in \mathcal{S}} \!\Bar{P}(s,a,s')\Bar{V}_{t+1}(s') \!\!\right)\!\!,
\end{equation*}
where $\bar P(s,a,s')$ denotes the privatized probability that taking action $a$ at state $s$ takes the agent to state $s'$.
We assume that the terminal values are given by a known function $R_T: \mathcal{S}\mapsto \mathbb{R}$, \textit{i.e.}, $\bar V_T(s) = R_T(s)$, for all $s\in \mathcal{S}$.\par
For an infinite-horizon discounted MDP, it can be shown that the optimal policy is a stationary policy, \textit{i.e.}, a policy that adopts the same decision rule at all stages \cite{c23}. Let $\bar V_\infty$ denote the optimal value of $\mathcal{\bar M}$. Then, for an infinite-horizon MDP, the second stage of Algorithm \ref{Alg:section3partA} computes $(\bar \pi, \bar V_\infty)$ such that for all $s \in \mathcal{S}$,
\begin{equation*} \label{eq:sec3subAeq3}
    \bar V_\infty (s) \!=\! \max_{\pi}\!\!\sum\limits_{a \in \mathcal{A}_s} \!\!\!\pi(a\!\mid\! s) \! \left(\!r(s,a) \!+\! \gamma\! \sum\limits_{s' \in \mathcal{S}} \!\Bar{P}(s,a,s')\Bar{V}_{\infty}(s') \!\!\right)\!\!,
\end{equation*}
\begin{equation*} \label{eq:sec3subAeq4}
    \bar\pi  \!\in\! \text{arg} \max\limits_\pi\!\!\sum\limits_{a \in \mathcal{A}_s} \!\!\!\pi(a\!\mid\! s) \! \left(\!r(s,a) \!+\! \gamma\! \sum\limits_{s' \in \mathcal{S}} \!\Bar{P}(s,a,s')\Bar{V}_{\infty}(s') \!\!\right)\!\!.
\end{equation*}
There are various methods suggested to efficiently compute $\bar \pi$ and its value function, such as dynamic programming or linear programming \cite{c23}.
The third and the fourth step of Algorithm \ref{Alg:section3partA} may adopt any of these methods to synthesize and evaluate an optimal policy for the privatized MDP $\mathcal{\bar M}$.

\subsection{Proof of differential privacy}
We prove that Algorithm \ref{Alg:section3partA} is $(\epsilon,\delta)$-differentially private by differential privacy's immunity to post-processing.
\begin{lemma}[From \cite{c6}, Proposition 2.1] \label{Lemma:postprocessing}
Let $\mathcal{M}: \Delta(n) \mapsto \Delta(n)$ be a mechanism that is $(\epsilon,\delta)$-differentially private. Let $f:\Delta(n)\mapsto \mathbb{R}$ be an arbitrary mapping. Then, $f \circ \mathcal{M} : \Delta(n) \mapsto \mathbb{R}$ is $(\epsilon, \delta)$-differentially private.
\end{lemma}
Recall that probabilistic $(\epsilon,\delta)$-differential privacy implies ordinary $(\epsilon,\delta)$-differential privacy.
Let $(\hat\epsilon,\hat\delta)$ denote the level of the probabilistic differential privacy of the Dirichlet mechanism employed in Algorithm \ref{Alg:section3partA}.
By Lemma \ref{Lemma:postprocessing}, the algorithm is $(\hat\epsilon,\hat\delta)$-differentially private because the synthesis step is an instance of a post-processing mapping $f$.

\section{utility analysis}\label{sec:costofprivacy}

Recall that Algorithm \ref{Alg:section3partA} synthesizes a policy $\bar\pi$ based on privatized transition probabilities in $\Bar{\mathcal{P}}$.
It then computes the value function of $\bar\pi$, $\bar V^{\bar \pi}_t$, using $\Bar{\mathcal{P}}$.
Let $V^{\bar \pi}_t : \mathcal{S}\mapsto\mathbb{R}$ be the value function that the non-private transition probabilities in $\mathcal{P}$ assign to $\bar \pi$.
The utility of Algorithm \ref{Alg:section3partA} is equal to $V^{\bar \pi}(s_0)$. \par

We assume that after the privatization stage, the algorithm loses access to the non-private transition probabilities in $\mathcal{P}$.
The reason is that in many real-world applications, a central cloud is used to compute the policy, and agents submit their data to the cloud \cite{c25, c26}.
For agents to preserve their data privacy, they privatize their data prior to any submission to the cloud \cite{c27}.

Had we had access to the non-private transition probabilities $\mathcal{P}$, $V^{\bar\pi}(s_0)$ could have been computed using an off-the-shelf policy evaluation algorithm.
We start off the utility analysis of Algorithm \ref{Alg:section3partA} with introducing a concentration bound on the output of the Dirichlet mechanism. 
\begin{lemma} \label{Lemma:concentration}
Let $\mathcal{M}^{(k)}_D$ denote a Dirichlet mechanism with parameter $k\in\mathbb{R}_+$. Then, for all $\beta > 0$, and all $p\in\Delta^\circ (n)$,
\begin{equation*}
    \mathbb{P}\left(\left\|\mathcal{M}_D^{(k)}(p)-p\right\|_\infty \geq \sqrt{\frac{\log\left(1/\beta\right)}{2(k+1)}}\right) \leq \beta.
\end{equation*}
\begin{proof}
See Appendix \ref{appendix: A}.
\end{proof}
\end{lemma}
The above lemma enables us to evaluate $V^{\bar\pi}(s_0)$, \textit{i.e.}, the conditional expectation of the value function without privacy, based on the privatized transition probabilities $\mathcal{\bar P}$ and $k$. 
In particular, for a finite-horizon MDP we provide an upper bound on $\left|\textbf{E}\left[V^{\bar \pi}_0(s_0)\mid\Bar{\mathcal{P}},k\right]-\Bar{V}^{\bar \pi}_0(s_0)\right|$. 
For an infinite-horizon MDP, we upper bound $\left|\textbf{E}\left[V^{\bar \pi}_\infty(s_0)\mid\Bar{\mathcal{P}},k\right]-\Bar{V}^{\bar \pi}_\infty(s_0)\right|$.
We refer to both expressions as the ``cost of privacy.''\par
The bounds are based on the pessimistic and optimistic value functions that possible transition-probability vectors generate.
Let $\alpha := \sqrt{{\log(1/\beta)}/{2(k+1)}}$, then Lemma \ref{Lemma:concentration} implies that for all $P(s,a)\in\mathcal{P}$ and the corresponding $\bar P(s,a)\in\mathcal{\bar P}$, $\mathbb{P}\left(||\bar P(s,a) - P(s,a)||_\infty \leq \alpha\right) \geq 1-\beta$. 
We define $\mathcal{\hat P}_{\alpha,\beta}$ as
\begin{multline}\label{eq:Phat}
     \{\beta P_1(s,a) \! + \! (1 \! - \! \beta) P_2(s,a) \mid \|P_2(s,a) - \Bar{P}(s,a)\|_\infty \leq \alpha,\\
     P_1(s,a),P_2(s,a) \in \Delta(|\mathcal{S}|), (s,a) \in \mathcal{S} \times \mathcal{A}_s\}
\end{multline}
We use the set $\hat{\mathcal{P}}_{\alpha,\beta}$ to compute a pessimistic and optimistic value function to bound the cost of privacy.
\subsection{Finite-horizon MDPs}
We bound the cost of privacy for a finite-horizon MDP by establishing a common upper and lower bound for both $\textbf{E}\left[V^{\bar \pi}_0(s_0)\mid\Bar{\mathcal{P}},k\right]$ and $\Bar{V}^{\bar \pi}_0(s_0)$.
We first state a technical lemma that we later use to prove the main theorem of this section.
\begin{lemma}\label{lemma:expectationwithinphat}
Fix $k$ and a set of transition probabilities $\mathcal{P}$, and let $\mathcal{\bar P}:=\left\{\bar P(s,a) = \mathcal{M}_D^{(k)}\left(P(s,a)\right) \mid P(s,a) \in \mathcal{P}\right\}$. For any $\beta>0$, let $\alpha := \sqrt{{\log(1/\beta)}/{2(k+1)}}$. Then, 
\begin{equation*}
    \bar P(s,a) \in \mathcal{\hat P}_{\alpha,\beta}, \ \forall\bar P(s,a) \in \mathcal{\bar P},
\end{equation*}
\begin{equation*}
    \mathbf{E}\left[P(s,a)\mid\mathcal{\bar P},k\right] \in \mathcal{\hat P}_{\alpha,\beta}, \ \forall P(s,a)\in\mathcal{P}.
\end{equation*}
\begin{proof}
See Appendix \ref{appendix: B}.
\end{proof}
\end{lemma}
\begin{theorem}\label{thrm: finitehorizon}
Let $\mathcal{M} = (\mathcal{S},\mathcal{A}_s, r, \mathcal{P}, T, \gamma)$ and $k$ be the input, and $(\bar\pi,\bar{V}_t^{\bar \pi})$ be the output of Algorithm \ref{Alg:section3partA}, and let $T<\infty$. Fix $\beta>0$, and let $\alpha := \sqrt{{\log(1/\beta)}/{2(k+1)}}$. Let $R_T:\mathcal{S}\mapsto\mathbb{R}$ denote the terminal value function of $\mathcal{M}$. Define $\ubar{v}^{\bar\pi}_t:\mathcal{S}\mapsto \mathbb{R}$ and $\Bar{v}^{\bar\pi}_t:\mathcal{S}\mapsto \mathbb{R}$ as follows. For all $s\in\mathcal{S}$, let $\bar{v}_T^{\bar \pi}(s) := R_T(s)$, $\ubar{v}_T^{\bar \pi}(s):=R_T(s)$, and for all $t\in\{0,\dots,T-1\}$, let
\begin{align*}
    &\ubar{v}^{\bar\pi}_t(s) \!\! := \!\!\sum\limits_{a \in \mathcal{A}_s} \! \! \bar\pi(a\!\mid\! s) \!\!\left(\! r(s,a)\! +\!\gamma\! \!\min\limits_{p \in \hat{\mathcal{P}}_{\alpha,\beta}} \! \sum\limits_{s' \in \mathcal{S}} p(s,a,s')\ubar{v}^{\bar\pi}_{t+1}(s') \!\right)\!\!,
\end{align*}
\begin{align*}
    &\bar{v}^{\bar\pi}_t(s) \!\! := \!\!\sum\limits_{a \in \mathcal{A}_s} \! \! \bar\pi(a\!\mid\! s) \!\!\left(\! r(s,a)\! +\!\gamma\! \!\max\limits_{p \in \hat{\mathcal{P}}_{\alpha,\beta}} \! \sum\limits_{s' \in \mathcal{S}} p(s,a,s')\bar{v}^{\bar\pi}_{t+1}(s') \!\right)\!\!.
\end{align*}
Then, we have that
\begin{equation*}
    \left|\mathbf{E}\left[V^{\bar\pi}_{0}(s_0)\mid\Bar{\mathcal{P}},k\right]-\Bar{V}_0^{\bar \pi}(s_0)\right| \leq \Bar{v}^{\bar\pi}_0(s_0) - \ubar{v}^{\bar\pi}_0(s_0).
\end{equation*}
\begin{proof}
We first show by induction that for all stages $t\in\{0,1,\dots,T\}$, and all states $s\in\mathcal{S}$, $\ubar{v}^{\bar\pi}_t(s)$ lower bounds $\Bar{V}_t^{\bar \pi}(s)$. 
Since all terminal values are determined by $R_T$, we have that for all states, ${\bar V}^{\bar \pi}_T(s) = \ubar{v}^{\bar \pi}_T(s) = R_T(s)$.
Let $\tau \in \{1,\dots,T-1\}$, and assume that for all states, we have that $\Bar{V}_\tau^{\bar \pi}(s) \geq \ubar{v}^{\bar\pi}_\tau(s)$. Then, for $t = \tau - 1$, and for all $s\in\mathcal{S}$, we can write
\begin{align}\label{eq:sec4partatheorem1eq1}
    \bar{V}^{\bar\pi}_{\tau-1}(s) \! &= \!\!\sum\limits_{a \in \mathcal{A}_s} \! \! \bar\pi(a\!\mid\! s) \!\!\left(\! r(s,a)\! +\!\gamma\! \! \sum\limits_{s' \in \mathcal{S}} \bar P(s,a,s')\bar{V}^{\bar\pi}_{\tau}(s') \!\right)\!\! \nonumber \\ 
    &\hspace{-0.4in} \geq \!\!\sum\limits_{a \in \mathcal{A}_s} \! \! \bar\pi(a\!\mid\! s) \!\!\left(\! r(s,a)\! +\!\gamma\! \! \sum\limits_{s' \in \mathcal{S}} \bar P(s,a,s')\ubar{v}^{\bar\pi}_{\tau}(s') \!\right)\!\!\nonumber \\ 
    &\hspace{-0.4in} \geq \!\!\sum\limits_{a \in \mathcal{A}_s} \! \! \bar\pi(a\!\mid\! s) \!\!\left(\! r(s,a)\! +\!\gamma\! \!\min\limits_{p \in \hat{\mathcal{P}}_{\alpha,\beta}} \! \sum\limits_{s' \in \mathcal{S}} p(s,a,s')\ubar{v}^{\bar\pi}_{\tau}(s') \!\right)\\
    &\hspace{-0.4in} = \ubar{v}^{\bar\pi}_{\tau-1}(s).\nonumber
\end{align}
The first inequality immediately results from the induction hypothesis.
For the second inequality, note that by Lemma \ref{lemma:expectationwithinphat}$, \Bar{P}(s,a) \in \hat{\mathcal{P}}$; therefore, $\Bar{P}(s,a)$ is a feasible solution of the minimization problem in \eqref{eq:sec4partatheorem1eq1}. \par
By induction, we conclude that for all $t\in\{0,\dots,T\}$, and all $s\in \mathcal{S}$, $\ubar{v}^{\bar\pi}_t(s)$ lower bounds $\Bar{V}_t^{\bar \pi}(s)$, which includes stage $t=0$ and state $s = s_0$, hence, $\Bar{V}_0^{\bar \pi}(s_0) \geq \ubar{v}^{\bar\pi}_0(s_0)$.
An identical argument by reversing the direction of the inequalities and substituting the minimization in \eqref{eq:sec4partatheorem1eq1} with maximization, leads to $\bar{v}^{\bar\pi}_t(s)$ upper bounding $\Bar{V}_t^{\bar \pi}(s)$, for all $s\in\mathcal{S}$ and all $t\in\{0,\dots,T\}$.
So far we have established that
\begin{equation} \label{eq:sec4partatheorem1eq2}
    \ubar{v}^{\bar\pi}_0(s_0) \leq \Bar{V}_0^{\bar \pi}(s_0) \leq \bar{v}^{\bar\pi}_0(s_0).
\end{equation}\par
We now consider the value of $\bar \pi$ based on the non-private transition probabilities in $\mathcal{P}$, that is denoted $V^{\bar{\pi}}$. 
Recall that for all stages $t\in\{1,\dots,T\}$ and all states $s\in\mathcal{S}$, $V^{\bar{\pi}}_t(s)$ satisfies the following conditions: $V_T^{\bar\pi} (s) = R_T(s)$ and
\begin{equation*}
    {V}^{\bar\pi}_{t-1}(s) \! = \!\!\sum\limits_{a \in \mathcal{A}_s} \! \! \bar\pi(a\!\mid\! s) \!\!\left(\! r(s,a)\! +\!\gamma\! \! \sum\limits_{s' \in \mathcal{S}} P(s,a,s')V^{\bar\pi}_{t}(s') \!\right)\!\!.
\end{equation*}
Taking the expectation of both sides, we arrive at
\begin{multline*}
    \mathbf{E}\left[{V}^{\bar\pi}_{t-1}(s)\,\middle\vert\,\mathcal{\bar P},k\right] = \\
    \sum\limits_{a \in \mathcal{A}_s} \! \! \bar\pi(a\!\mid\! s) \!\!\left(\! r(s,a)\! +\!\gamma\! \! \sum\limits_{s' \in \mathcal{S}} \mathbf{E}\left[P(s,a,s')V^{\bar\pi}_{t}(s')\,\middle\vert\,\mathcal{\bar P},k\right] \!\right)\!\!.
\end{multline*} \par
In order to to establish an iterative relation between the values of $\mathbf{E}\left[{V}^{\bar\pi}_{t-1}(s)\,\middle\vert\,\mathcal{\bar P},k\right]$ for different stages, we need to break the expectation on the right-hand side of the above equation.
Consider the scenario wherein at each stage, the transition probabilities are independently privatized using the Dirichlet mechanism. Let $\Tilde{V}^{\bar\pi}$ denote the value function that is associated with $\bar\pi$ according to the above scenario. Due to the independence assumption, we can write
\begin{multline*}
    \mathbf{E}\left[\tilde V^{\bar\pi}_{t-1}(s)\,\middle\vert\,\mathcal{\bar P},k\right] =\left. \sum\limits_{a \in \mathcal{A}_s} \! \! \bar\pi(a\!\mid\! s) \right(\! r(s,a) + \\
    \left.\!\gamma\! \! \sum\limits_{s' \in \mathcal{S}} \mathbf{E}\left[P(s,a,s')\,\middle\vert\,\mathcal{\bar P},k\right] \mathbf{E}\left[\tilde V^{\bar\pi}_{t}(s')\,\middle\vert\,\mathcal{\bar P},k\right] \!\right)\!\!.
\end{multline*}\par
Assume that for an arbitrary $\tau\in\{1,\dots,T-1\}$ and for all $s\in\mathcal{S}$, it holds that $\mathbf{E}\left[\tilde V^{\bar\pi}_{\tau}(s')\mid\mathcal{\bar P},k\right] \geq \ubar{v}_\tau^{\bar\pi}(s)$. Then, for $t = \tau - 1$, it holds that
\begin{multline*}
    \mathbf{E}\left[{\tilde V}^{\bar\pi}_{\tau-1}(s)\,\middle\vert\,\mathcal{\bar P},k\right] \geq \\
    \sum\limits_{a \in \mathcal{A}_s} \! \! \bar\pi(a\!\mid\! s) \!\!\left(\! r(s,a)\! +\!\gamma\! \! \sum\limits_{s' \in \mathcal{S}} \mathbf{E}\left[P(s,a,s')\,\middle\vert\,\mathcal{\bar P},k\right]\ubar{v}^{\bar\pi}_{\tau}(s') \!\right)\!\!.
\end{multline*}\par
By Lemma \ref{lemma:expectationwithinphat}, $\mathbf{E}\left[P(s,a)\mid\mathcal{\bar P},k\right]$ belongs to the set $\mathcal{\hat{P}}_{\alpha,\beta}$.
Therefore we can write
\begin{multline*}
    \mathbf{E}\left[{\tilde V}^{\bar\pi}_{\tau-1}(s)\mid\mathcal{\bar P},k\right] \geq \\
    \sum\limits_{a \in \mathcal{A}_s} \! \! \!\bar\pi(a\!\mid\! s) \!\!\left(\! \!r(s,a)\! +\!\gamma\! \! \sum\limits_{s' \in \mathcal{S}} \min\limits_{p\in\mathcal{\hat P}_{\alpha,\beta}} \!\!\! p(s,a,s')\ubar{v}^{\bar\pi}_{\tau}(s') \! \!\right)\!\!\! = \! \ubar{v}^{\bar\pi}_{\tau-1}(s).
\end{multline*}
Therefore, by induction, we have shown that for all stages and all states, $\ubar{v}^{\bar\pi}$ is a lower bound to $\mathbf{E}\left[{\tilde V}^{\bar\pi}\mid\mathcal{\bar P},k\right]$. \par
We now revisit the independence assumption.
We introduce the worst-case expected value function that possible transition probabilities can generate as the lower bound to the value function of the generated policy without privacy. 
The independence assumption leads to independent minimization problems at each stage.
The worst-case value function without the independence assumption has the following restriction on the transition probabilities.
Transition probabilities that appear as a minimization variable for the same state-action pair at different stages must take equal values.
Since the feasible region of the case without independence is a subset of that of the case with independence, the lower bound for $\mathbf{E}\left[\tilde V^{\bar\pi}\mid\mathcal{\bar P},k\right]$ also lower bounds $\mathbf{E}\left[{V}^{\bar\pi}\mid\mathcal{\bar P},k\right]$.\par
Similarly, an identical argument by reversing the direction of the inequalities, and substituting the min operators with max operators leads to $\mathbf{E}\left[{\tilde V}^{\bar\pi}\mid\mathcal{\bar P},k\right] \leq \bar{v}^{\bar\pi}$, for all stages and all states.
Therefore,
\begin{equation} \label{eq:sec4partatheorem1eq5}
    \ubar{v}^{\bar\pi}_0(s_0) \leq \mathbf{E}\left[{V}^{\bar\pi}_{0}(s_0)\mid\mathcal{\bar P},k\right] \leq \bar{v}^{\bar\pi}_0(s_0).
\end{equation}
Combining \eqref{eq:sec4partatheorem1eq2} and \eqref{eq:sec4partatheorem1eq5} concludes the proof.
\end{proof}
\end{theorem}
The assumption of independence has also been adopted in \cite{c15,c16,c17} wherein the problem of robust MDPs is studied. 
In the above works, the scenario wherein the independence assumption holds is called the dynamic model and the counterpart scenario is called the static model.
See Section 2.2 of \cite{c15} for further details about the relationship between the static and the dynamic model.\par
\subsection{Infinite-horizon MDPs}
In this section, we bound the cost of privacy for an infinite-horizon MDP.
We first state a technical lemma, which we later use to bound the cost of privacy for infinite-horizon MDPs.

\begin{lemma}\label{lemma:mappings}
Fix $k$ and an MDP $\mathcal{M} = (\mathcal{S},\mathcal{A}_s, r, \mathcal{P}, T, \gamma)$, and let $\mathcal{\bar P}:=\left\{\bar P(s,a) = \mathcal{M}_D^{(k)}\left(P(s,a)\right) \mid P(s,a) \in \mathcal{P}\right\}$. For any $\beta>0$, let $\alpha := \sqrt{{\log(1/\beta)}/{2(k+1)}}$. Define mappings $\mathcal{L}_1,\mathcal{L}_2,\mathcal{L}_3: \mathbb{R}^{|\mathcal{S}|} \times \mathbb{R}^{|\mathcal{S}|}$ as
\begin{align}
    &\mathcal{L}_1 \mathbf{v}\! := \!\!\sum\limits_{a \in \mathcal{A}_s} \! \! \bar\pi(a\!\mid\! s) \!\!\left(\! r(s,a)\! +\!\gamma\! \! \min\limits_{p\in\mathcal{\hat P}_{\alpha,\beta}}\sum\limits_{s' \in \mathcal{S}}  p(s,a,s')\mathbf{v}(s') \!\right)\!\!, \nonumber\\
    &\mathcal{L}_2 \mathbf{v}\! :=\! \! \sum\limits_{a \in \mathcal{A}_s} \! \! \bar\pi(a\!\mid\! s) \!\!\left(\! r(s,a)\! +\!\gamma\! \! \sum\limits_{s' \in \mathcal{S}} \bar P(s,a,s')\mathbf{v}(s') \!\right)\!\!\nonumber,\\
    &\mathcal{L}_3 \mathbf{v}\! := \!\!\sum\limits_{a \in \mathcal{A}_s} \! \! \bar\pi(a\!\mid\! s) \!\!\left(\! r(s,a)\! +\!\gamma\! \! \sum\limits_{s' \in \mathcal{S}} \!\! \mathbf{E}\left[P(s,a,s')\mathbf{v}(s')\!\,\middle\vert\,\!\mathcal{\bar P},k\right] \!\!\right)\!\!. \nonumber
\end{align}
Then, mappings $\mathcal{L}_1$, $\mathcal{L}_2$, and $\mathcal{L}_3$ are $\gamma$-contraction mappings, \textit{i.e.}, for all $\mathbf{v_1}, \mathbf{v_2} \in\mathbb{R}^{|\mathcal{S}|}$ and $i\in\{1,2,3\}$, 
\begin{equation*}
    \left\|\mathcal{L}_i \mathbf{v_1} - \mathcal{L}_i \mathbf{v_2}\right\|_\infty \leq \gamma \left\| \mathbf{v_1} - \mathbf{v_2} \right\|_\infty.
\end{equation*}
\begin{proof}
See Appendix \ref{appendix: C}.
\end{proof}
\end{lemma}

\begin{theorem} \label{thrm: infinitehorizon}
Let $\mathcal{M} = (\mathcal{S},\mathcal{A}_s, r, \mathcal{P}, T, \gamma)$ and $k$ be the input, and $(\bar\pi,\bar{V}_\infty^{\bar \pi})$ be the output of Algorithm \ref{Alg:section3partA}, and let $T=\infty$. Fix $\beta>0$, and let $\alpha := \sqrt{{\log(1/\beta)}/{2(k+1)}}$. For all $s\in\mathcal{S}$, let $\ubar{v}^{\bar\pi}_\infty:\mathcal{S}\mapsto \mathbb{R}$ and $\bar{v}^{\bar\pi}_\infty:\mathcal{S}\mapsto \mathbb{R}$ satisfy
\begin{align*}
    &\ubar{v}^{\bar\pi}_\infty(s) \!\! = \!\!\sum\limits_{a \in \mathcal{A}_s} \! \! \bar\pi(a\!\mid\! s) \!\!\left(\! r(s,a)\! +\!\gamma\! \!\min\limits_{p \in \hat{\mathcal{P}}_{\alpha,\beta}} \! \sum\limits_{s' \in \mathcal{S}} p(s,a,s')\ubar{v}^{\bar\pi}_{\infty}(s') \!\right)\!\!,
\end{align*}
\begin{align*}
    &\bar{v}^{\bar\pi}_\infty(s) \!\! = \!\!\sum\limits_{a \in \mathcal{A}_s} \! \! \bar\pi(a\!\mid\! s) \!\!\left(\! r(s,a)\! +\!\gamma\! \!\max\limits_{p \in \hat{\mathcal{P}}_{\alpha,\beta}} \! \sum\limits_{s' \in \mathcal{S}} p(s,a,s')\bar{v}^{\bar\pi}_{\infty}(s') \!\right)\!\!.
\end{align*}
Then, we have that
\begin{equation*}
    \left|\mathbf{E}\left[V^{\bar\pi}_{\infty}(s_0)\mid\Bar{\mathcal{P}},k\right]-\Bar{V}_\infty^{\bar \pi}(s_0)\right| \leq \Bar{v}^{\bar\pi}_\infty(s_0) - \ubar{v}^{\bar\pi}_\infty(s_0).
\end{equation*}
\begin{proof}
Consider the $\gamma$-contraction mappings $\mathcal{L}_1$, $\mathcal{L}_2$ and $\mathcal{L}_3$ introduced in Lemma \ref{lemma:mappings}.
Notice that $\ubar v ^{\bar \pi}_\infty$, $\bar V^{\bar\pi}\infty$ and $\mathbf{E}\left[V^{\bar\pi}_{\infty}(s_0)\mid\Bar{\mathcal{P}},k\right]$ are the fixed points of mappings $\mathcal{L}_1$, $\mathcal{L}_2$ and $\mathcal{L}_3$, respectively.
Contraction mappings are known to admit a fixed point by Banach fixed-point theorem.
Furthermore, contractive mappings converge to their fixed points by applying the mapping repeatedly to an arbitrary initial $v\in\mathbb{R}^{|\mathcal{S}|}$.
Let $v_i^{(k)}$ denote the $k^{\text{th}}$ iteration corresponding to $\mathcal{L}_i$, and let all three mappings start from a common initial vector.
Assume that at stage $k$, it holds that $v_1^{(k)} \le v_2^{(k)}$, then for stage $k+1$, with a similar argument to the proof of Theorem \ref{thrm: finitehorizon}, we can write
\begin{align}\label{eq:sec4partbtheorem2eq1}
    v_2^{(k+1)}(s) \! &= \!\!\sum\limits_{a \in \mathcal{A}_s} \! \! \bar\pi(a\!\mid\! s) \!\!\left(\! r(s,a)\! +\!\gamma\! \! \sum\limits_{s' \in \mathcal{S}} \bar P(s,a,s')v_2^{(k)}(s') \!\right)\!\! \nonumber \\ 
    &\hspace{-0.4in} \geq \!\!\sum\limits_{a \in \mathcal{A}_s} \! \! \bar\pi(a\!\mid\! s) \!\!\left(\! r(s,a)\! +\!\gamma\! \! \sum\limits_{s' \in \mathcal{S}} \bar P(s,a,s')v_1^{(k)}(s') \!\right)\!\!\nonumber \\ 
    &\hspace{-0.4in} \geq \!\!\sum\limits_{a \in \mathcal{A}_s} \! \! \bar\pi(a\!\mid\! s) \!\!\left(\! r(s,a)\! +\!\gamma\! \!\min\limits_{p \in \hat{\mathcal{P}}_{\alpha,\beta}} \! \sum\limits_{s' \in \mathcal{S}} p(s,a,s')v_1^{(k)}(s') \!\right) \nonumber\\
    &\hspace{-0.4in} = v_1^{(k+1)}(s).\nonumber
\end{align}
As a result, we have that the limiting value of $v_1^{k}$ and $v_2^{k}$ when $k\to\infty$ also satisfy
\begin{equation*}
   \ubar v_\infty^{\bar\pi} = \lim\limits_{k\to\infty} v_1^{(k)} \le \lim\limits_{k\to\infty} v_2^{(k)} = \bar V^{\bar \pi}_\infty.
\end{equation*}
Using a similar argument, it can be shown that $\bar V^{\bar \pi}_\infty$ is upper bounded by $\bar v_\infty^{\bar\pi}$.
Therefore we have that
\begin{equation}\label{eq:sec4partbtheorem2eq2}
    \ubar v_\infty^{\bar\pi} \le \bar V^{\bar \pi}_\infty \le \bar v_\infty^{\bar\pi}.
\end{equation}
We now assume that at stage $k$, it holds that $v_1^{(k)} \le v_3^{(k)}$. At stage $k+1$, we write
\begin{multline*}
    v_3^{(k+1)}(s)= \\
    \sum\limits_{a \in \mathcal{A}_s} \! \! \bar\pi(a\!\mid\! s) \!\!\left(\! r(s,a)\! +\!\gamma\! \! \sum\limits_{s' \in \mathcal{S}}\!\! \mathbf{E}\left[P(s,a,s')v_3^{(k)}(s')\,\middle\vert\,\mathcal{\bar P},k\right] \!\right)\!\!.
\end{multline*}
Notice that there is no stochasticity in $v_3^{(k)}$. Therefore, we have that
\begin{multline*}
    v_3^{(k+1)}(s)= \\
    \sum\limits_{a \in \mathcal{A}_s} \! \! \bar\pi(a\!\mid\! s) \!\!\left(\! r(s,a)\! +\!\gamma\! \! \sum\limits_{s' \in \mathcal{S}}\!\! \mathbf{E}\left[P(s,a,s')\,\middle\vert\,\mathcal{\bar P},k\right]v_3^{(k)}(s') \!\right)\!\!.
\end{multline*}
The above equation combined with the induction hypothesis implies that  
\begin{multline*}
    v_3^{(k+1)}(s)\ge \\
    \sum\limits_{a \in \mathcal{A}_s} \! \! \bar\pi(a\!\mid\! s) \!\!\left(\! r(s,a)\! +\!\gamma\! \! \sum\limits_{s' \in \mathcal{S}}\!\! \mathbf{E}\left[P(s,a,s')\,\middle\vert\,\mathcal{\bar P},k\right]v_1^{(k)}(s') \!\right)\!\!.
\end{multline*}
By Lemma \ref{lemma:expectationwithinphat}, we have that $v_3^{(k+1)} \ge v_3^{(k+1)}$. Therefore,
\begin{equation}\label{eq:sec4partbtheorem2eq3}
   \ubar v_\infty^{\bar\pi} = \lim\limits_{k\to\infty} v_1^{(k)} \le \lim\limits_{k\to\infty} v_3^{(k)} = \mathbf{E}\left[V^{\bar\pi}_{\infty}(s_0)\mid\Bar{\mathcal{P}},k\right].
\end{equation}
With a similar argument, it can be shown that $\bar v_\infty^{\bar \pi}$ upper bounds $\mathbf{E}\left[V^{\bar\pi}_{\infty}(s_0)\mid\Bar{\mathcal{P}},k\right]$, which combined by \eqref{eq:sec4partbtheorem2eq2} and \eqref{eq:sec4partbtheorem2eq3} concludes the proof.
\end{proof}
\end{theorem}

\begin{figure*}[t]
    \input{plots}
    \caption{Plots a and b show all the value functions that are used to compute and validate the cost of privacy for Examples 1 and 2. The third plot shows the cost of privacy itself for both examples.\label{fig:results}}
\end{figure*}
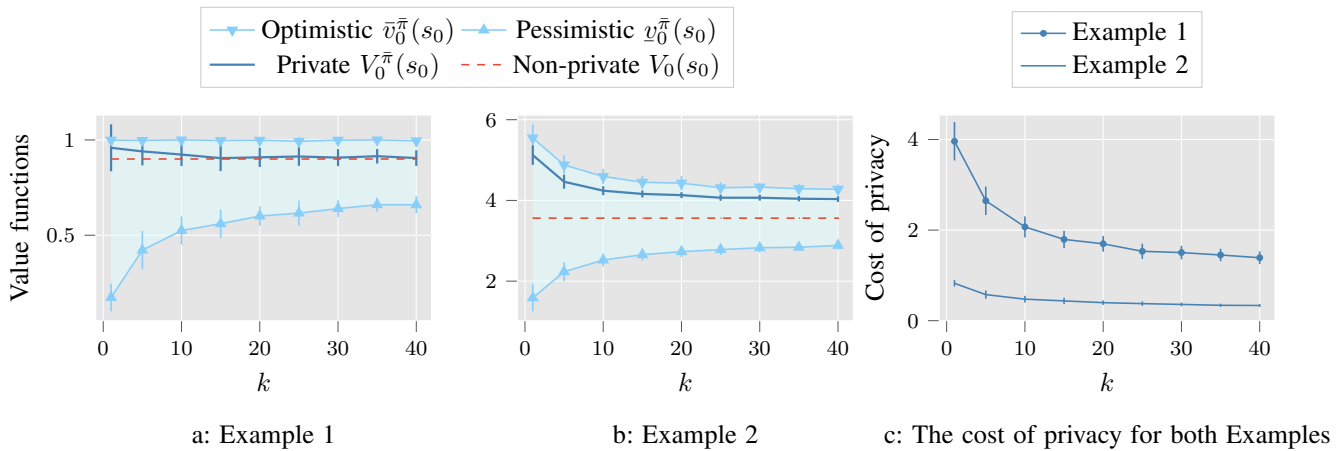
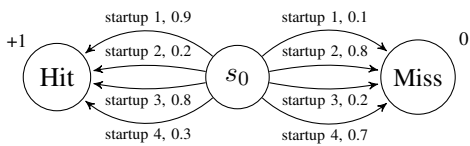
\begin{figure}
\centering
\input{mdp}
\caption{The corporation's investment model with four possible startups to acquire, given as an MDP.\label{Fig: investmentMDP}}
\end{figure}

\section{computational complexity} \label{sec:compcomplexity}
In the previous section, we introduced expressions that bound the cost of privacy for both finite- and infinite-horizon MDPs.
The bounds do not take a closed form, and they are computed by iterative methods.
In this section, we show that the cost of privacy for both cases can be computed efficiently.
In particular we show that for both cases, the computational complexity is polynomial in problem parameters.
\subsection{Finite-horizon MDPs}
Revisiting the definition of $\ubar{v}^{\bar\pi}_t$ and $\bar{v}^{\bar\pi}_t$ in Theorem \ref{thrm: finitehorizon}, the inner minimization or maximization problem must be solved for each of $T$ stages and $|\mathcal{S}|$ states.
We first consider computing the lower bound $\ubar{v}^{\bar\pi}_t$.
Fix $(s,a)\in\mathcal{S}\times\mathcal{A}_s$ and $t\in\{0,\dots,T-1\}$.
Then the inner minimization problem can be recast as
\begin{align}\label{OP:inner_optimization_finite_horizon}\tag{P}
    &\min\limits_{P_1, P_2, p \in \mathbb{R}^{|\mathcal{S}|}} & &\sum\limits_{s'\in\mathcal{S}} p(s,a,s')\ubar{v}^{\bar\pi}_t(s')& &\\
    &\text{subject to} & &\mathbf{1}^T P_1(s,a) = 1, P_1(s,a,s')\ge0, & &\!\!\!\!\!\!\forall s'\in\mathcal{S},\nonumber\\
    & & &\mathbf{1}^T P_2(s,a) = 1,P_2(s,a,s')\ge0, & &\!\!\!\!\!\!\forall s'\in\mathcal{S},\nonumber\\
    & & &\mathbf{1}^T p(s,a) = 1, p(s,a,s') \ge0, & &\!\!\!\!\!\!\forall s'\in\mathcal{S},\nonumber\\
    & & &P_2(s,a,s') - \bar P(s,a,s') \leq \alpha, & &\!\!\!\!\!\!\forall s'\in\mathcal{S},\nonumber\\
    & & &P_2(s,a,s') - \bar P(s,a,s') \geq -\alpha, & &\!\!\!\!\!\!\forall s'\in\mathcal{S},\nonumber\\
    & & &\beta P_1(s,a) + (1-\beta) P_2(s,a) = p(s,a). & &\nonumber
\end{align}
The above optimization problem is a linear program (LP) with $3|\mathcal{S}|$ variables and $6|\mathcal{S}|+3$ constraints. 
Similarly, the inner maximization problem in $\bar{v}^{\bar\pi}_t$ can be cast as an LP by negating the objective function in \eqref{OP:inner_optimization_finite_horizon}. \par
There exists numerous algorithms for solving an LP, each of which is associated with different computational complexities.
We consider the interior-point method that is known to solve an LP in polynomial time, $\mathcal{O}(n^{3.5})$, where $n$ is the number of variables \cite{karmarkar1984new}.
Therefore, the computational complexity of computing the cost of privacy for a finite-horizon MDP is $\mathcal{O}\left(T|\mathcal{S}|^{4.5} |\mathcal{A}_s|\right)$.

\subsection{Infinite-horizon MDPs}
in Lemma \ref{lemma:mappings}, we introduced $\mathcal{L}_1$ that is a $\gamma$-contraction mapping. 
Suppose there exists a constant $R\in\mathbb{R}_+$ such that the reward function of the underlying MDP satisfies $|r(s,a)|\le R$, for all $(s,a)\in\mathcal{S}\times\mathcal{A}_s$.
Then, all the value functions including the private, non-private, optimistic, and the pessimistic value function must be bounded above by a constant $v_\text{max}$.
Let $v_1^{(k)}$ be the value of the $k^\text{th}$ iteration corresponding to $\mathcal{L}_1$, and $v_1^{\infty}$ be the limiting value.
We can write
\begin{equation*}
    \left\| v_1^{(k+1)} - v_1^{\infty} \right\|_\infty \leq \gamma \left\| v_1^{(k)} - v_1^{\infty} \right\|_\infty.
\end{equation*}
The above inequality is equivalent to
\begin{equation*}
    \left\| v_1^{(k)} - v_1^{\infty} \right\|_\infty \leq \frac{\gamma^k}{1-\gamma} \left\| v_1^{(1)} - v_1^{(0)} \right\|_\infty \leq 2v_\text{max}\frac{\gamma^k}{1-\gamma}.
\end{equation*}
The above inequality indicates that in order to reach an $\epsilon$-approximation of the limit, $\mathcal{O}(\log(1/\epsilon))$ iterations are required.
The inner minimization problem is identical to the finite-horizon case, which we reformulated as an LP in \eqref{OP:inner_optimization_finite_horizon}.
Combining the above arguments together, we conclude that the required number of iterations such that $\|v_1^{(k)}-\ubar{v}_\infty^{\bar\pi}\|_\infty \le \epsilon$, is $\mathcal{O}\left(|\mathcal{S}|^{4.5} |\mathcal{A}_s| \log\left(1/\epsilon\right)\right)$.
The same computational complexity holds for the upper bound $\bar v^{\bar\pi}_\infty$.
As a result, the cost of privacy for infinite-horizon MDPs can be computed in polynomial time as well.

\section{numerical results} \label{sec:numericalresults}
In this section, we empirically validate the developments of previous sections, wherein we introduced the expressions that compute the cost of privacy and their corresponding computational complexity.
We apply Algorithm \ref{Alg:section3partA}  to two example MDPs at a range of $k$ values, which represent a range of privacy protection levels. 
The first is a small-sized MDP that represents a simple model for a corporation's investment planning.
The second example is an MDP with a larger state and action space, which has transition probabilities, reward function, and terminal reward function generated randomly. 
For both examples, the algorithm is run 50 times, and Figure \ref{fig:results}, which depicts the results, shows the mean values alongside their standard deviations that appear as error bars. \par

\textbf{Example 1.} Suppose a corporation has been tracking four startups, and it has to decide which startup to acquire. Assume that the corporation's model of each of the startup's probability of success is given by the MDP in Figure \ref{Fig: investmentMDP}.\par

The first empirical result of this section corresponds to applying the algorithm to the scenario described in Example 1, and is depicted in Figure \ref{fig:results}a.
The results show that as $k$ increases, the pessimistic and optimistic value functions provide better bounds for the private and non-private value functions.
Therefore the cost of privacy decreases with $k$, which Figure \ref{fig:results}c confirms. \par

\textbf{Example 2.} 
In this example, we apply the algorithm to a larger MDP in order to test its scalability.
In particular, the MDP has 20 states, 5 actions available at each state, and a time horizon of 10.\par
Similar to the previous example, Figure \ref{fig:results}b indicates that an increase in $k$ improves the approximations of the private and non-private value functions by $\bar v^{\bar \pi}_0(s_0)$ and $\ubar v^{\bar \pi}_0(s_0)$.
As a result, the cost of privacy must decrease with $k$, which is confirmed by Figure \ref{fig:results}c.\par

In Example 2, the computation of the cost of privacy for each instance of $k$ took 4.88s on a desktop computer with a 3.5GHz CPU and 24GB of RAM.
Doubling the time horizon of the MDP to 20 results in 9.34s/itr, and doubling the size of the action space to $|\mathcal{A}_s| = 10$ results in 9.68s/itr, which is consistent with the computational complexity introduced in Section \ref{sec:compcomplexity}.

For both examples, the negative correlation between $k$ and the cost of privacy is consistent with the developments in Section \ref{sec:costofprivacy}.
By Lemma \ref{Lemma:concentration}, an increase in $k$ results in a tighter concentration bound on the output of the Dirichlet mechanism, and it lowers $\alpha$ in Theorems \ref{thrm: finitehorizon} and \ref{thrm: infinitehorizon}.
A smaller $\alpha$ further restricts the inner optimization problem in \eqref{OP:inner_optimization_finite_horizon}; thus, it helps the optimistic and the pessimistic value functions to provide better approximations, which leads to a lower cost of privacy.

\section{conclusion}\label{sec:conclusion}
We introduced a privacy-preserving policy synthesis algorithm that protects the privacy of the transition probabilities of its input MDP.
The algorithm employs a Dirichlet mechanism to privatize the transition probabilities.
We established a concentration bound on the output of the Dirichlet mechanism based on its scaling parameter $k$.
We used the concentration bound to bound the cost of privacy imposed by privatizing the transition probabilities.
We further showed that the cost of privacy can be computed efficiently by establishing that the computational complexity of the algorithm is polynomial in problem parameters.
Finally, the simulation results validated the developments in both the soundness of the expressions we introduced for the cost of privacy and the computational complexity associated with computing them.\par
\bibliographystyle{IEEEtran}

\appendices
\section{Proof of Lemma \ref{Lemma:concentration}} \label{appendix: A}
Let $p\in\Delta^\circ(n)$ and $k>0$. From the properties of the Dirichlet mechanism we have that $\mathbf{E}\left[\mathcal{M}_D^{(k)}(p)\right] = p$.
Let $u\in\Delta(n)$ and $\lambda\in \mathbb{R}_+$. By Theorem 3.3 in \cite{c24}, we have that
\begin{multline*} \label{eq:appendixAeq1}
    \textbf{E}\left[\exp{\left(\lambda u^T\left(\mathcal{M}_D^{(k)}(p)-\textbf{E}\left[\mathcal{M}_D^{(k)}(p)\right]\right)\right)}\right] \leq\\ \prod\limits_{i=1}^n \textbf{E}\left[\exp{\left(\lambda u_i \left(\mathcal{M}_D^{(k)}(p)_i - p_i\right)\right)}\right].
\end{multline*}
Let $\text{beta}(a,b)$ denote the beta distribution with parameter $(a,b)$. It can be shown that each component of the Dirichlet mechanism satisfies
\begin{equation*}
    \mathcal{M}_D^{(k)}(p)_i \sim \text{beta}(kp_i,k(1-p_i)).
\end{equation*}
Recall that a random variable $X$ is said to be $R$-subgaussian if it satisfies 
\begin{equation*}
    \mathbf{E}\left[\exp{\left(sX\right)}\right] \leq \exp{\left(\frac{R^2 s^2}{2}\right)}, \forall s\in\mathbb{R}.
\end{equation*}
A beta distribution with parameters $(a,b)$ is $\sqrt{1/4(a+b+1)}$-subgaussian \cite{c24}.
Therefore,
\begin{multline*}
    \prod\limits_{i=1}^n \textbf{E}\left[\exp{\left(\lambda u_i \left(\mathcal{M}_D^{(k)}(p)_i - p_i\right)\right)}\right] \leq\\ 
     \prod\limits_{i=1}^d \exp{\left(\frac{\lambda^2 u_i^2 \sigma_i^2}{2}\right)} = \exp{\left(\frac{\lambda^2 \|u\|_2}{8(k+1)}\right)}.
\end{multline*}
Since $u$ can be any vector in $\Delta(n)$, we consider an instance of $u$ that puts weight $1$ on the component of $\mathcal{M}_D^{(k)}(p)-p$ with maximum magnitude and zero elsewhere. Considering the case where $\lambda = 1$, we can write
\begin{equation}\label{eq:expectationbounnd}
    \textbf{E}\left[\exp{\left\|\mathcal{M}_D^{(k)}(p)-p\right\|_\infty}\right] \leq \exp{\left(\frac{1}{8(k+1)}\right)}.
\end{equation}
Finally, for all $\theta\geq0$, we have that
\begin{multline*}
     \mathbb{P}\left(\left\|\mathcal{M}_D^{(k)}(p)-p\right\|_\infty \geq \alpha\right) =\\ \mathbb{P}\left(\exp\left( \theta\left\|\mathcal{M}_D^{(k)}(p)-p\right\|_\infty\right) \geq \exp\left(\theta\alpha\right)\right).
\end{multline*}
By Markov's inequality,
\begin{multline*}
    \mathbb{P}\left(\exp\left( \theta\left\|\mathcal{M}_D^{(k)}(p)-p\right\|_\infty\right) \geq \exp\left(\theta\alpha\right)\right) \leq \\ \textbf{E}\left[\exp\left( \theta\left\|\mathcal{M}_D^{(k)}(p)-p\right\|_\infty\right)\right] \cdot\exp\left(-\theta\alpha\right).
\end{multline*}
Combining the above inequality with \eqref{eq:expectationbounnd}, we arrive at
\begin{equation*}
     \mathbb{P}\left(\left\|\mathcal{M}_D^{(k)}(p)-p\right\|_\infty \geq \alpha\right)\leq \exp\left(\frac{\theta^2}{8(k+1)} - \theta \alpha\right).
\end{equation*}
Taking $\theta = 4\alpha (k+1)$ and a proper change of variable from $\alpha$ to $\beta$ concludes the lemma.

\section{Proof of Lemma \ref{lemma:expectationwithinphat}} \label{appendix: B}
In the definition of $\mathcal{\hat P}_{\alpha,\beta}$ in \eqref{eq:Phat}, take $P_1(s,a) := \Bar{P}(s,a)$, and $P_2(s,a) := \Bar{P}(s,a)$.
Then, $\beta P_1(s,a) + (1-\beta) P_2(s,a) = \bar P(s,a)$ is within the set $\mathcal{\hat P}_{\alpha,\beta}$, which concludes the first result.

For a given $\beta$, by Lemma \ref{Lemma:concentration}, there exists $\beta' \geq \beta$, such that
\begin{equation*}
    \mathbb{P}\left(\left\|\mathcal{M}_D^{(k)}(p)-p\right\|_\infty \geq \sqrt{\frac{\log\left(1/\beta'\right)}{2(k+1)}}\right) = \beta.
\end{equation*}
Define $\alpha':=\sqrt{{\log(1/\beta')}/{2(k+1)}}$ and let
\begin{equation*}
    P_1(s,a) := \textbf{E}\left[P(s,a) \,\middle\vert\, \mathcal{\bar P},k,\|P(s,a) - \Bar{P}(s,a)\|_\infty \geq \alpha' \right],
\end{equation*}
and
\begin{equation*}
    P_2(s,a) := \textbf{E}\left[P(s,a) \,\middle\vert\, \mathcal{\bar P},k,\|P(s,a) - \Bar{P}(s,a)\|_\infty \leq \alpha' \right].
\end{equation*}
Since $\alpha' \leq \alpha$, $\|P_2(s,a) - \Bar{P}(s,a)\|_\infty \leq \alpha$. 
Then, $\beta P_1(s,a) + (1-\beta) P_2(s,a) = \mathbf{E}\left[P(s,a)\mid\mathcal{\bar P},k\right]$ is also within the set $\mathcal{\hat P}_{\alpha,\beta}$, which completes the proof.

\section{Proof of Lemma \ref{lemma:expectationwithinphat}} \label{appendix: C}
Let $U, V \in \mathbb{R}^{|\mathcal{S}|}$. Then, for all $s\in\mathcal{S}$, we can write
\begin{multline*}
        \mathcal{L}_1 U(s) - \mathcal{L}_1 V(s)  =\!\gamma\! \!\!\sum\limits_{a \in \mathcal{A}_s} \! \! \bar\pi(a\!\mid\! s)\cdot  \\
        \!\!\left(\!\min\limits_{p\in\mathcal{\hat P}_{\alpha,\beta}} \!\!\sum\limits_{s' \in \mathcal{S}} \right. p(s,a,s')U(s')-\!\!\left.\min\limits_{p\in\mathcal{\hat P}_{\alpha,\beta}}\!\!\sum\limits_{s' \in \mathcal{S}} p(s,a,s')V(s')\!\right).
\end{multline*}
For all $(s,a)\in\mathcal{S}\times\mathcal{A}_s$, let 
\begin{equation*}
    p^*(s,a) = \mathop{\mathrm{argmin}} \limits_{p\in\mathcal{\hat P}_{\alpha,\beta}}\!\!\sum\limits_{s' \in \mathcal{S}} p(s,a,s')V(s').
\end{equation*} 
Then,
\begin{multline}\label{eq:appendixIIIeq1}
    \mathcal{L}_1 U(s) - \mathcal{L}_1 V(s) \le \gamma\! \!\!\sum\limits_{a \in \mathcal{A}_s}\bar\pi(a\!\mid\! s)\cdot\\\sum\limits_{s'\in\mathcal{S}} p^*(s,a,s')(U(s')-V(s'))\!\!
     \le \!\!\gamma\max\limits_{s\in\mathcal{S}}\left|U(s)-V(s)\right|.
\end{multline}
For mapping $\mathcal{L}_2$, we have that
\begin{multline}\label{eq:appendixIIIeq2}
        \mathcal{L}_2 U(s) - \mathcal{L}_2 V(s)  =\!\gamma\! \!\!\sum\limits_{a \in \mathcal{A}_s} \! \! \bar\pi(a\!\mid\! s)\cdot  \\
        \!\!\left(\sum\limits_{s' \in \mathcal{S}} \!\! \bar P(s,a,s')\!\left(U(s')\!-\!V(s')\right)\!\!\right)\!\le\! \gamma\max\limits_{s\in\mathcal{S}}\left|U(s)\!-\!V(s)\right|.
\end{multline}
Finally, for $\mathcal{L}_3$, we write
\begin{multline}\label{eq:appendixIIIeq3}
        \mathcal{L}_2 U(s) \!-\! \mathcal{L}_2 V(s)  =\!\gamma\! \!\!\sum\limits_{a \in \mathcal{A}_s} \! \! \bar\pi(a\!\mid\! s)\!\!\sum\limits_{s' \in \mathcal{S}} \!\!\mathbf{E}\left[P(s,a,s')\,\middle\vert\,\mathcal{\bar P},k\right]\cdot  \\
        \left(U(s')\!-\!V(s')\right)\!\le\! \gamma\max\limits_{s\in\mathcal{S}}\left|U(s)\!-\!V(s)\right|\!.
\end{multline}
Notice that $\left\|U-V\right\|_\infty = \max\limits_{s\in\mathcal{S}}\left|U(s)\!-\!V(s)\right|$, and that \eqref{eq:appendixIIIeq1}, \eqref{eq:appendixIIIeq2} and \eqref{eq:appendixIIIeq3} hold for any $s\in\mathcal{S}$. Thus, $\mathcal{L}_1$, $\mathcal{L}_2$ and $\mathcal{L}_3$ are $\gamma$-contraction mappings.
\end{document}

%% file: plots.tex
\begin{tikzpicture}

\definecolor{color0}{rgb}{0.529411764705882,0.807843137254902,0.980392156862745}
\definecolor{color1}{rgb}{0.87843137254902,1,1}
\definecolor{color2}{rgb}{0.274509803921569,0.509803921568627,0.705882352941177}
\definecolor{color3}{rgb}{0.886274509803922,0.290196078431373,0.2}

\begin{groupplot}[group style={group name = my plots, group size=3 by 1, horizontal sep = 0.45in}, height = 1.7in, width = 0.34\textwidth]
\nextgroupplot[
axis background/.style={fill=white!89.8039215686275!black},
axis line style={white},
legend cell align={left},
legend style={fill opacity=0.8, draw opacity=1, text opacity=1, at={(0.97,0.03)}, anchor=south east, draw=white!80!black, fill=white!89.8039215686275!black},
tick align=outside,
tick pos=left,
x grid style={white},
xlabel={\(\displaystyle k\)},
ylabel = {Value functions},
xmajorgrids,
xmin=-0.95, xmax=41.95,
xtick style={color=white!33.3333333333333!black},
y grid style={white},
ymajorgrids,
ymin=0.0516464111358635, ymax=1.13075461756078,
ytick style={color=white!33.3333333333333!black}
]
\path [fill=color1, fill opacity=0.5, very thin]
(axis cs:1,0.9988453981822)
--(axis cs:1,0.9988453981822)
--(axis cs:5,0.997543628166654)
--(axis cs:10,1)
--(axis cs:15,0.997241574710609)
--(axis cs:20,0.998537412065669)
--(axis cs:25,0.992961107014797)
--(axis cs:30,0.998787271563482)
--(axis cs:35,1)
--(axis cs:40,0.995295718265407)
--(axis cs:40,0.995295718265407)
--(axis cs:40,0.659791189094576)
--(axis cs:35,0.660198457311902)
--(axis cs:30,0.639601793195036)
--(axis cs:25,0.61605116891086)
--(axis cs:20,0.600523060707289)
--(axis cs:15,0.560526832044294)
--(axis cs:10,0.524716549118633)
--(axis cs:5,0.42185742004777)
--(axis cs:1,0.17329085294322)
--cycle;

\path [draw=color0, semithick]
(axis cs:1,0.993321170870778)
--(axis cs:1,1.00436962549362);

\path [draw=color0, semithick]
(axis cs:5,0.980694552692895)
--(axis cs:5,1.01439270364041);

\path [draw=color0, semithick]
(axis cs:10,1)
--(axis cs:10,1);

\path [draw=color0, semithick]
(axis cs:15,0.985561016153003)
--(axis cs:15,1.00892213326822);

\path [draw=color0, semithick]
(axis cs:20,0.99078301741602)
--(axis cs:20,1.00629180671532);

\path [draw=color0, semithick]
(axis cs:25,0.972634865846381)
--(axis cs:25,1.01328734818321);

\path [draw=color0, semithick]
(axis cs:30,0.993225976899292)
--(axis cs:30,1.00434856622767);

\path [draw=color0, semithick]
(axis cs:35,1)
--(axis cs:35,1);

\path [draw=color0, semithick]
(axis cs:40,0.980394385077483)
--(axis cs:40,1.01019705145333);

\path [draw=color0, semithick]
(axis cs:1,0.100696784155178)
--(axis cs:1,0.245884921731262);

\path [draw=color0, semithick]
(axis cs:5,0.320944557956293)
--(axis cs:5,0.522770282139246);

\path [draw=color0, semithick]
(axis cs:10,0.450904096626152)
--(axis cs:10,0.598529001611114);

\path [draw=color0, semithick]
(axis cs:15,0.486654035960951)
--(axis cs:15,0.634399628127637);

\path [draw=color0, semithick]
(axis cs:20,0.549653208542071)
--(axis cs:20,0.651392912872506);

\path [draw=color0, semithick]
(axis cs:25,0.550507523430825)
--(axis cs:25,0.681594814390894);

\path [draw=color0, semithick]
(axis cs:30,0.595576583382062)
--(axis cs:30,0.68362700300801);

\path [draw=color0, semithick]
(axis cs:35,0.623674698579258)
--(axis cs:35,0.696722216044547);

\path [draw=color0, semithick]
(axis cs:40,0.61554729654149)
--(axis cs:40,0.704035081647661);

\path [draw=color2, line width=0.84pt]
(axis cs:1,0.835800407028053)
--(axis cs:1,1.08170424454146);

\path [draw=color2, line width=0.84pt]
(axis cs:5,0.86622162215267)
--(axis cs:5,1.01361585595448);

\path [draw=color2, line width=0.84pt]
(axis cs:10,0.863373215149659)
--(axis cs:10,0.983047630945558);

\path [draw=color2, line width=0.84pt]
(axis cs:15,0.836822031376878)
--(axis cs:15,0.972238633711973);

\path [draw=color2, line width=0.84pt]
(axis cs:20,0.858382243439217)
--(axis cs:20,0.959143402646618);

\path [draw=color2, line width=0.84pt]
(axis cs:25,0.864255724091751)
--(axis cs:25,0.961987214828185);

\path [draw=color2, line width=0.84pt]
(axis cs:30,0.862454497066882)
--(axis cs:30,0.951835469130038);

\path [draw=color2, line width=0.84pt]
(axis cs:35,0.876542561163196)
--(axis cs:35,0.952792517092064);

\path [draw=color2, line width=0.84pt]
(axis cs:40,0.86383562302142)
--(axis cs:40,0.945898235166577);

\path [draw=color3, semithick]
(axis cs:1,0.9)
--(axis cs:1,0.9);

\path [draw=color3, semithick]
(axis cs:5,0.9)
--(axis cs:5,0.9);

\path [draw=color3, semithick]
(axis cs:10,0.9)
--(axis cs:10,0.9);

\path [draw=color3, semithick]
(axis cs:15,0.9)
--(axis cs:15,0.9);

\path [draw=color3, semithick]
(axis cs:20,0.9)
--(axis cs:20,0.9);

\path [draw=color3, semithick]
(axis cs:25,0.9)
--(axis cs:25,0.9);

\path [draw=color3, semithick]
(axis cs:30,0.9)
--(axis cs:30,0.9);

\path [draw=color3, semithick]
(axis cs:35,0.9)
--(axis cs:35,0.9);

\path [draw=color3, semithick]
(axis cs:40,0.9)
--(axis cs:40,0.9);

\addplot [semithick, color0, mark=triangle*, mark size=2, mark options={solid, rotate = 180}]
table {%
1 0.9988453981822
5 0.997543628166654
10 1
15 0.997241574710609
20 0.998537412065669
25 0.992961107014797
30 0.998787271563482
35 1
40 0.995295718265407
};
\addplot [semithick, color0, mark=triangle*, mark size=2, mark options={solid}]
table {%
1 0.17329085294322
5 0.42185742004777
10 0.524716549118633
15 0.560526832044294
20 0.600523060707289
25 0.61605116891086
30 0.639601793195036
35 0.660198457311902
40 0.659791189094576
};
\addplot [line width=0.84pt, color2]
table {%
1 0.958752325784758
5 0.939918739053577
10 0.923210423047609
15 0.904530332544425
20 0.908762823042918
25 0.913121469459968
30 0.90714498309846
35 0.91466753912763
40 0.904866929093999
};
\addplot [semithick, color3, dashed]
table {%
1 0.9
5 0.9
10 0.9
15 0.9
20 0.9
25 0.9
30 0.9
35 0.9
40 0.9
};

\nextgroupplot[
axis background/.style={fill=white!89.8039215686275!black},
axis line style={white},
legend columns=2,
legend style={at={(-0.15,1.132)}, anchor=south, draw=white!80.0!black, fill=white!100!black},
tick align=outside,
tick pos=left,
x grid style={white},
xlabel={\(\displaystyle k\)},
xmajorgrids,
xmin=-0.95, xmax=41.95,
xtick style={color=white!33.3333333333333!black},
y grid style={white},
ymajorgrids,
ymin=1.02036430721041, ymax=6.11624827448566,
ytick style={color=white!33.3333333333333!black}
]
\path [fill=color1, fill opacity=0.5, very thin]
(axis cs:1,5.54727087119407)
--(axis cs:1,5.54727087119407)
--(axis cs:5,4.87992669688925)
--(axis cs:10,4.5936891569023)
--(axis cs:15,4.44997315778454)
--(axis cs:20,4.42829317719793)
--(axis cs:25,4.31381242671069)
--(axis cs:30,4.33101439184814)
--(axis cs:35,4.29079864738827)
--(axis cs:40,4.27622753073866)
--(axis cs:40,4.27622753073866)
--(axis cs:40,2.88616502367316)
--(axis cs:35,2.8401228502656)
--(axis cs:30,2.82703361684993)
--(axis cs:25,2.78201357388944)
--(axis cs:20,2.73090395406758)
--(axis cs:15,2.65537502029012)
--(axis cs:10,2.52304019855641)
--(axis cs:5,2.23260877043967)
--(axis cs:1,1.58668123651491)
--cycle;

\path [draw=color0, semithick]
(axis cs:1,5.20992455732408)
--(axis cs:1,5.88461718506406);

\path [draw=color0, semithick]
(axis cs:5,4.64108710544737)
--(axis cs:5,5.11876628833112);

\path [draw=color0, semithick]
(axis cs:10,4.40609272434135)
--(axis cs:10,4.78128558946326);

\path [draw=color0, semithick]
(axis cs:15,4.29211927311989)
--(axis cs:15,4.60782704244919);

\path [draw=color0, semithick]
(axis cs:20,4.25014983229254)
--(axis cs:20,4.60643652210331);

\path [draw=color0, semithick]
(axis cs:25,4.18059438617886)
--(axis cs:25,4.44703046724253);

\path [draw=color0, semithick]
(axis cs:30,4.22353632870813)
--(axis cs:30,4.43849245498816);

\path [draw=color0, semithick]
(axis cs:35,4.17189484761287)
--(axis cs:35,4.40970244716366);

\path [draw=color0, semithick]
(axis cs:40,4.15552885544054)
--(axis cs:40,4.39692620603677);

\path [draw=color0, semithick]
(axis cs:1,1.25199539663201)
--(axis cs:1,1.9213670763978);

\path [draw=color0, semithick]
(axis cs:5,1.99806533197164)
--(axis cs:5,2.4671522089077);

\path [draw=color0, semithick]
(axis cs:10,2.36779693303141)
--(axis cs:10,2.67828346408141);

\path [draw=color0, semithick]
(axis cs:15,2.51064618052745)
--(axis cs:15,2.80010386005279);

\path [draw=color0, semithick]
(axis cs:20,2.59554840897178)
--(axis cs:20,2.86625949916337);

\path [draw=color0, semithick]
(axis cs:25,2.6494535198989)
--(axis cs:25,2.91457362787998);

\path [draw=color0, semithick]
(axis cs:30,2.71046411690541)
--(axis cs:30,2.94360311679445);

\path [draw=color0, semithick]
(axis cs:35,2.76022400464571)
--(axis cs:35,2.92002169588548);

\path [draw=color0, semithick]
(axis cs:40,2.78614870861362)
--(axis cs:40,2.98618133873269);

\path [draw=color2, line width=0.84pt]
(axis cs:1,4.88193355946472)
--(axis cs:1,5.36482456665861);

\path [draw=color2, line width=0.84pt]
(axis cs:5,4.29201226700958)
--(axis cs:5,4.63278124302311);

\path [draw=color2, line width=0.84pt]
(axis cs:10,4.13168065130691)
--(axis cs:10,4.35122743030235);

\path [draw=color2, line width=0.84pt]
(axis cs:15,4.07428934608168)
--(axis cs:15,4.24818916687208);

\path [draw=color2, line width=0.84pt]
(axis cs:20,4.05779676759444)
--(axis cs:20,4.20431185206026);

\path [draw=color2, line width=0.84pt]
(axis cs:25,3.99086246612159)
--(axis cs:25,4.14396812490355);

\path [draw=color2, line width=0.84pt]
(axis cs:30,3.99462304078323)
--(axis cs:30,4.1396396559131);

\path [draw=color2, line width=0.84pt]
(axis cs:35,3.97568704707964)
--(axis cs:35,4.10923254882727);

\path [draw=color2, line width=0.84pt]
(axis cs:40,3.96289781370315)
--(axis cs:40,4.10254421415015);

\path [draw=color3, semithick]
(axis cs:1,3.56064333656573)
--(axis cs:1,3.56064333656573);

\path [draw=color3, semithick]
(axis cs:5,3.56064333656573)
--(axis cs:5,3.56064333656573);

\path [draw=color3, semithick]
(axis cs:10,3.56064333656573)
--(axis cs:10,3.56064333656573);

\path [draw=color3, semithick]
(axis cs:15,3.56064333656573)
--(axis cs:15,3.56064333656573);

\path [draw=color3, semithick]
(axis cs:20,3.56064333656573)
--(axis cs:20,3.56064333656573);

\path [draw=color3, semithick]
(axis cs:25,3.56064333656573)
--(axis cs:25,3.56064333656573);

\path [draw=color3, semithick]
(axis cs:30,3.56064333656573)
--(axis cs:30,3.56064333656573);

\path [draw=color3, semithick]
(axis cs:35,3.56064333656573)
--(axis cs:35,3.56064333656573);

\path [draw=color3, semithick]
(axis cs:40,3.56064333656573)
--(axis cs:40,3.56064333656573);

\addplot [semithick, color0, mark=triangle*, mark size=2, mark options={solid, rotate = 180}]
table {%
1 5.54727087119407
5 4.87992669688925
10 4.5936891569023
15 4.44997315778454
20 4.42829317719793
25 4.31381242671069
30 4.33101439184814
35 4.29079864738827
40 4.27622753073866
};
\addlegendentry{Optimistic $\bar v^{\bar \pi}_0(s_0)$}
\addplot [semithick, color0, mark=triangle*, mark size=2, mark options={solid}]
table {%
1 1.58668123651491
5 2.23260877043967
10 2.52304019855641
15 2.65537502029012
20 2.73090395406758
25 2.78201357388944
30 2.82703361684993
35 2.8401228502656
40 2.88616502367316
};
\addlegendentry{Pessimistic $\ubar v^{\bar \pi}_0(s_0)$}
\addplot [line width=0.84pt, color2]
table {%
1 5.12337906306166
5 4.46239675501635
10 4.24145404080463
15 4.16123925647688
20 4.13105430982735
25 4.06741529551257
30 4.06713134834817
35 4.04245979795346
40 4.03272101392665
};
\addlegendentry{Private $V^{\bar \pi}_0(s_0)$}
\addplot [semithick, color3, dashed]
table {%
1 3.56064333656573
5 3.56064333656573
10 3.56064333656573
15 3.56064333656573
20 3.56064333656573
25 3.56064333656573
30 3.56064333656573
35 3.56064333656573
40 3.56064333656573
};
\addlegendentry{Non-private $V_0(s_0)$}

\nextgroupplot[
legend columns=1,
legend style={at={(0.5,1.132)}, anchor=south, draw=white!80.0!black, fill=white!100!black},
axis background/.style={fill=white!89.8039215686275!black},
axis line style={white},
tick align=outside,
tick pos=left,
x grid style={white},
xlabel={\(\displaystyle k\)},
ylabel={Cost of privacy},
xmajorgrids,
xmin=-0.95, xmax=41.95,
xtick style={color=white!33.3333333333333!black},
y grid style={white},
ymajorgrids,
ymin=0, ymax=4.54078229189229,
ytick style={color=white!33.3333333333333!black}
]
\path [draw=color2, semithick]
(axis cs:1,3.53707886623886)
--(axis cs:1,4.38410040311947);

\path [draw=color2, semithick]
(axis cs:5,2.33237108888468)
--(axis cs:5,2.96226476401448);

\path [draw=color2, semithick]
(axis cs:10,1.83825169966645)
--(axis cs:10,2.30304621702534);

\path [draw=color2, semithick]
(axis cs:15,1.60452198462517)
--(axis cs:15,1.98467429036367);

\path [draw=color2, semithick]
(axis cs:20,1.52398816697595)
--(axis cs:20,1.87079027928475);

\path [draw=color2, semithick]
(axis cs:25,1.36411303583401)
--(axis cs:25,1.69948466980849);

\path [draw=color2, semithick]
(axis cs:30,1.3550403144729)
--(axis cs:30,1.65292123552352);

\path [draw=color2, semithick]
(axis cs:35,1.31306394836345)
--(axis cs:35,1.58828764588189);

\path [draw=color2, semithick]
(axis cs:40,1.25046262766315)
--(axis cs:40,1.52966238646785);

\path [draw=color2, semithick]
(axis cs:1,0.755366069649601)
--(axis cs:1,0.895743020828358);

\path [draw=color2, semithick]
(axis cs:5,0.483617924562377)
--(axis cs:5,0.667754491675392);

\path [draw=color2, semithick]
(axis cs:10,0.401470998388887)
--(axis cs:10,0.549095903373848);

\path [draw=color2, semithick]
(axis cs:15,0.368625315630435)
--(axis cs:15,0.504804169702196);

\path [draw=color2, semithick]
(axis cs:20,0.349894284310268)
--(axis cs:20,0.446134418406494);

\path [draw=color2, semithick]
(axis cs:25,0.326373610827421)
--(axis cs:25,0.427446265380454);

\path [draw=color2, semithick]
(axis cs:30,0.317940453855798)
--(axis cs:30,0.400430502881094);

\path [draw=color2, semithick]
(axis cs:35,0.303277783955453)
--(axis cs:35,0.376325301420742);

\path [draw=color2, semithick]
(axis cs:40,0.301856073516851)
--(axis cs:40,0.369152984824812);

\addplot [semithick, color2, mark = *, mark size = 1]
table {%
1 3.96058963467917
5 2.64731792644958
10 2.07064895834589
15 1.79459813749442
20 1.69738922313035
25 1.53179885282125
30 1.50398077499821
35 1.45067579712267
40 1.3900625070655
};
\addlegendentry{Example 1}
\addplot [semithick, color2]
table {%
1 0.825554545238979
5 0.575686208118884
10 0.475283450881368
15 0.436714742666315
20 0.398014351358381
25 0.376909938103938
30 0.359185478368446
35 0.339801542688098
40 0.335504529170831
};
\addlegendentry{Example 2}
\end{groupplot}
\node[text width=6cm,align=center,anchor=north] at ([yshift=-10mm]my plots c1r1.south) {\captionof{subfigure}{Example 1}};
\node[text width=6cm,align=center,anchor=north] at ([yshift=-10mm]my plots c2r1.south) {\captionof{subfigure}{Example 2}};
\node[text width=6cm,align=center,anchor=north] at ([yshift=-10mm]my plots c3r1.south) {\captionof{subfigure}{The cost of privacy for both Examples}};
\end{tikzpicture}

%% file: mdp.tex
\begin{tikzpicture}[->,>=stealth',shorten >=1pt,auto,node distance=4cm,
        scale = 0.6,transform shape]

  \node[state] (q0) [font=\fontsize{11}{0},scale = 1.6] {$s_0$};
  \node[state, label = {above left:\large\hspace{-0.5cm}\text{+1}}] (q1) [left of=q0, font=\fontsize{5}{5}, scale = 1.7] {Hit};
  \node[state, label = {above right:\large\hspace{0.2cm}\text{0}}] (q2) [right of=q0, font=\fontsize{5}{5}, scale = 1.6] {Miss};

    \path (q0) edge [bend right=40] node [ above] {startup 1, 0.9} (q1);
    \path (q0) edge [bend right=10] node [ above] {startup 2, 0.2} (q1);
    \path (q0) edge [bend left=10] node [ below] {startup 3, 0.8} (q1);
    \path (q0) edge [bend left=40] node [ below] {startup 4, 0.3} (q1);

    \path (q0) edge [bend left=40] node [ above] {startup 1, 0.1} (q2);
    \path (q0) edge [bend left=10] node [ above] {startup 2, 0.8} (q2);
    \path (q0) edge [bend right=10] node [ below] {startup 3, 0.2} (q2);
    \path (q0) edge [bend right=40] node [ below] {startup 4, 0.7} (q2);

\end{tikzpicture}